\documentclass[
]{article}
\usepackage[top=20truemm,bottom=20truemm,left=20truemm,right=20truemm]{geometry}
\usepackage{authblk}
\usepackage{abstract}
\usepackage{amsmath}
\usepackage{amssymb}
\usepackage{hyperref}
\usepackage[dvipdfmx]{graphicx}
\usepackage{color}

\author[1]{Takuya Tsuchiya\thanks{\href{mailto:t-tsuchiya@hi-tech.ac.jp}{\nolinkurl{t-tsuchiya@hi-tech.ac.jp}}}}
\author[2]{Ryosuke Urakawa}
\author[3]{Gen Yoneda}
\affil[1]{%
  Center for Liberal Arts and Sciences,
  Hachinohe Institute of Technology,
  Japan
}
\affil[2]{%
  Waseda University Advanced Research Institute for Science and Engineering,
  Japan
}
\affil[3]{%
  Graduate School of Fundamental Science and Engineering, Waseda University,
  Japan
}

\title{Stable numerical simulation of Einstein equations
  \\in gravitational collapse space--time}

\begin{document}

\maketitle

\begin{abstract}
  We perform simulations in a gravitational collapsing model using the Einstein
  equations.
  In this paper, we review the equations for constructing the initial values
  and the evolution form of the Einstein equations called the BSSN formulation.
  In addition, since we treat a nonvacuum case, we review the evolution
  equations of the matter fields of a perfect fluid.
  To make the simulations stable, we propose a modified system, which decreases
  numerical errors in analysis,
  and we actually perform stable simulations with decreased numerical errors.
\end{abstract}

\section{Introduction}

Numerical relativity, which solves the Einstein equations numerically,
has been widely studied \cite{Alcubierre, Baumgarte-Shapiro, Gourgoulhon,
  Shibata}.
In particular, for the direct observation of gravitational waves
(e.g., \cite{Abbott16}), numerical relativity makes important contributions.

Stable numerical simulations are important to clarify the details of phenomena.
Since the Einstein equations are the nonlinear partial differential equations,
numerical errors tend to accumulate.
Although there are some research studies to reduce the numerical errors, they
are mainly in the vacuum case.
Thus, we suggest a modified system to reduce numerical errors by modifying the
evolution equation in a perfect fluid.

The structure of this paper is as follows.
We review the space--time decomposition of the Einstein equations in a perfect
fluid in Sec. \ref{sec:abst}.
In Sec. \ref{sec:equation}, we introduce the equations for the numerical
simulations of the Einstein equations as the initial value problem in the
perfect fluid.
We perform some simulations in the dust case and propose a modified system to
reduce the numerical errors in Sec. \ref{sec:simulation}.
We summarize this paper in Sec. \ref{sec:summary}.
In this paper, indices such as $(\mu, \nu, \lambda, \cdots)$ and
$(i, j, k, \cdots)$ run from 0 to 3 and 1 to 3, respectively.
We use the Einstein convention of summation of repeated up--down indices.

\section{Abstract for numerical simulations of Einstein equations in perfect
  fluid
  \label{sec:abst}
}

The Einstein equations are as follows.
\begin{align}
  {}^{(4)}R_{\mu\nu} - \dfrac{1}{2}{}^{(4)}Rg_{\mu\nu} = 8\pi T_{\mu\nu},
  \label{eq:Ein}
\end{align}
where $g_{\mu\nu}$ is the four-dimensional metric, ${}^{(4)}R_{\mu\nu}$ is the
four-dimensional Ricci tensor, ${}^{(4)}R\equiv g^{\mu\nu}{}^{(4)}R_{\mu\nu}$,
$g^{\mu\nu}$ is the inverse of $g_{\mu\nu}$, and $T_{\mu\nu}$ is the
stress--energy tensor.
Eq. \eqref{eq:Ein} is not a dynamical form because the time and space components
are mixed.
Generally, we often carry out space--time decomposition of Eq. \eqref{eq:Ein}
for performing the simulations.

There are some dynamical forms of Eq. \eqref{eq:Ein}, the most basic
formulation is the ADM formulation
\cite{Arnowitte-Deser-Misner08, Smarr-York78}:
\begin{align}
  \partial_t\gamma_{ij}
  &= -2\alpha K_{ij} + \gamma_{jm}(D_i\beta^m) + \gamma_{im}(D_j\beta^m),
  \label{eq:gamma_ADM}\\
  \partial_tK_{ij}
  &= - D_iD_j\alpha + \alpha R_{ij} + \alpha KK_{ij}
  -2\alpha \gamma^{mn}K_{mi}K_{nj}
  \nonumber\\
  &\quad
  -8\pi \alpha S_{ij} + 4\pi \alpha S\gamma_{ij}
  -4\pi \alpha\rho_{\text{H}}\gamma_{ij}
  + \beta^\ell(D_\ell K_{ij})
  \nonumber\\
  &\quad
  + K_{\ell i}(D_j\beta^\ell)
  + K_{\ell j}(D_i\beta^\ell),
  \label{eq:K_ADM}\\
  \mathcal{H}
  &\equiv R + K^2 - \gamma^{im}\gamma^{nj}K_{ij} K_{mn}
  - 16\pi \rho_{\text{H}}\approx 0,
  \label{eq:H_ADM}\\
  \mathcal{M}_i
  &\equiv \gamma^{jm}(D_jK_{mi}) - D_i K - 8\pi J_i\approx0,
  \label{eq:M_ADM}
\end{align}
where $\gamma_{ij}\equiv g_{ij}$ is the induced metric, $\gamma^{ij}$ is the
inverse of $\gamma_{ij}$, and $K_{ij}$ is the extrinsic curvature defined as
Eq. \eqref{eq:gamma_ADM}.
$\alpha\equiv1/\sqrt{-g^{00}}$ and $\beta^i\equiv\alpha^2g^{0i}$ are the lapse
function and the shift vector, respectively.
$D_i$ is the covariant derivative associated with $\gamma_{ij}$, $R_{ij}$ is
the Ricci tensor in three dimensions, $K\equiv\gamma^{ij}K_{ij}$, and
$R\equiv\gamma^{ij}R_{ij}$.
$\rho_{\mathrm{H}}\equiv\alpha^2 g^{0\mu}g^{0\nu}T_{\mu\nu}$ is the mass
density, $J_i\equiv\alpha g^{0\mu} T_{\mu i}$ is the momentum density,
$S_{ij}\equiv T_{ij}$ is the stress tensor, and $S\equiv\gamma^{ij}S_{ij}$.
$\mathcal{H}$ and $\mathcal{M}_i$ are the Hamiltonian constraint and the
momentum constraint, respectively.
The symbol $\approx$ means zero in the mathematical sense but nonzero in the
numerical sense.

In the calculation in the nonvacuum case, we also solve the evolution equations
of matter fields.
These equations are given by the space--time decomposition of
$g^{\mu\lambda}(\nabla_\mu T_{\lambda\nu})=0$, where
$\nabla_\mu$ is the covariant derivative associated with $g_{\mu\nu}$.
For the perfect fluid case, the stress--energy tensor is given as
\begin{align}
  T_{\mu\nu} = \{\rho(1+\epsilon)+p\}g_{\mu\lambda}g_{\nu\omega}
  u^\lambda u^\omega + pg_{\mu\nu},
\end{align}
where $\rho$ is the rest mass, $\epsilon$ is the inner energy, $u^\mu$ is the
four velocity, and $p$ is the pressure.
The evolution equations of $\epsilon$ and $\rho u^i$ are given by
$g^{\mu\lambda}(\nabla_\mu T_{\lambda\nu})=0$.
The evolution equation of $\rho$ is given by the continuous equation
$\nabla_\mu (\rho u^\mu)=0$.
On the other hand, $p$ is usually given by other conditions such as the
equation of state.

\section{Basic equations
  \label{sec:equation}
}

\subsection{Equations for initial values}

If we set the initial values, they have to satisfy the constraints
$\mathcal{H}$ and $\mathcal{M}_i$.
The dynamical variables $\gamma_{ij}$ and $K_{ij}$ include 12 components because
they are symmetric tensors.
However, there are only four constraints.
Thus, according to \cite{OMurchadha-York74}, we reformulate the constraints such
that
\begin{align}
  \hat{\gamma}^{ij}(\hat{D}_i\hat{D}_j\psi)
  &= \frac{1}{8}\psi\hat{R}
  + \frac{1}{12}\psi^{5}K^2
  - 2\pi\psi^{5}\rho_{\text{H}}
  \nonumber\\
  &\quad
  - \frac{1}{8\psi^{7}}\hat{\gamma}^{im}\hat{\gamma}^{jn}\hat{A}_{ij}
  \hat{A}_{mn},
  \label{eq:ini_psi}\\
  \hat{\gamma}^{jm}(\hat{D}_j\hat{D}_mX_i)
  &=
  - \frac{1}{3}\hat{\gamma}^{jm}(\hat{D}_i\hat{D}_jX_m)
  + \frac{2}{3}\psi^{6}(\hat{D}_iK)
  \nonumber\\
  &\quad
  - \hat{R}_{ij}\hat{\gamma}^{jm}X_m
  + 8\pi \psi^{6}J_i,
  \label{eq:ini_X}
\end{align}
where $\hat{A}_{ij}\equiv \psi^{2}(K_{ij} - (1/3)K\psi^4\hat{\gamma}_{ij})$,
$\psi$ satisfies relations such as $\gamma_{ij}=\psi^{4}\hat{\gamma}_{ij}$, and
$X_i$ satisfies the relation $\hat{D}_iX_j + \hat{D}_jX_i
- (2/3)\hat{\gamma}^{mn}(\hat{D}_n X_m)\hat{\gamma}_{ij}= \hat{A}_{ij}$.
$\hat{D}_i$ is the covariant derivative associated with $\hat{\gamma}_{ij}$,
$\hat{R}_{ij}$ is the Ricci tensor of $\hat{\gamma}^{ij}$, and
$\hat{R}\equiv \hat{\gamma}^{ij}\hat{R}_{ij}$.
We often assume $\hat{\gamma}_{ij}=\delta_{ij}$ and $K=0$.
Then, we set a suitable boundary condition, and we solve Eqs.
\eqref{eq:ini_psi}--\eqref{eq:ini_X}.

The Einstein equations are not satisfied by simply giving the dynamical
variables $(\gamma_{ij}, K_{ij})$ at the initial time.
We have to give the gauge variables $(\alpha, \beta^i)$ as appropriate values.
For the lapse function $\alpha$, we assume $K=0$ and $\partial_tK=0$ in Eqs.
\eqref{eq:gamma_ADM}--\eqref{eq:H_ADM}, and we obtain
\begin{align}
  \hat{\gamma}^{ij}(\hat{D}_i\hat{D}_j\alpha)
  &= -2\psi^{-1}\hat{\gamma}^{ij}(\hat{D}_j\psi)(\hat{D}_i\alpha)
  + 4\pi \alpha \psi^{4}\rho_{\text{H}}
  \nonumber\\
  &\quad
  + \alpha \psi^{-8}\hat{\gamma}^{mi}\hat{\gamma}^{jn}\hat{A}_{ij}\hat{A}_{mn}
  + 4\pi \psi^4\alpha S.
  \label{eq:maximal}
\end{align}
This is called the maximal slicing condition
\cite{Estabrook-Wahlquist-Christensen-DeWitte-Smarr-Tsiang73}.
For the shift vector $\beta^i$, the minimal distortion gauge condition
\cite{Smarr-York78} is often used.
We solve the nonlinear elliptic-type Eqs.
\eqref{eq:ini_psi}--\eqref{eq:maximal},
obtain $(\psi, \hat{X}_i,\alpha)$, and then get $(\alpha, \gamma_{ij}, K_{ij})$
at the initial time.

\subsection{Evolution equations and constraint equations
  \label{sec:BSSN}
}

Since it is well known that the numerical simulations using
Eqs. \eqref{eq:gamma_ADM}--\eqref{eq:K_ADM} are unstable, we should reformulate
the evolution equations.
The BSSN formulation \cite{Shibata-Nakamura95, Baumgarte-Shapiro98} is one of
the formulations most commonly used by numerical relativists.
The dynamical variables are $(\varphi, K, \tilde{\gamma}_{ij}, \tilde{A}_{ij},
\tilde{\Gamma}^i)$ instead of $(\gamma_{ij}, K_{ij})$, where
$\varphi=(1/12)\log(\text{det}(\gamma_{ij}))$,
$\tilde{\gamma}_{ij}=e^{-4\varphi}\gamma_{ij}$, $K=\gamma^{ij}K_{ij}$,
$\tilde{A}_{ij}=e^{-4\varphi}(K_{ij}-(1/3)K\gamma_{ij})$,
$\tilde{\Gamma}^\ell=\tilde{\Gamma}^\ell{}_{ij}\tilde{\gamma}^{ij}$,
$\tilde{\gamma}^{ij}$ is the inverse of $\tilde{\gamma}_{ij}$, and
$\tilde{\Gamma}^\ell{}_{ij}$ is the connection coefficient of
$\tilde{\gamma}_{ij}$.
The evolution equations are
\begin{align}
  \partial_t \varphi
  &=
  - \frac{1}{6}\alpha K
  + \frac{1}{6}(\partial_i \beta^i)
  + \beta^i (\partial_i \varphi),
  \label{eq:varphiEvo}\\
  \partial_t K
  &=
  - (D_i  D_j \alpha)e^{-4\varphi}\tilde{\gamma}^{ij}
  + \frac{1}{3}\alpha K^2
  + 4\pi\alpha S
  \nonumber\\
  &\quad
  + \alpha\tilde{A}_{ij}\tilde{A}_{mn}\tilde{\gamma}^{im}\tilde{\gamma}^{jn}
  + 4\pi\alpha\rho_{\text{H}}
  + \beta^i (\partial_i  K),
  \label{eq:KEvo}\\
  \partial_t\tilde{\gamma}_{i j }
  &=
  - 2\alpha\tilde{A}_{i j }
  - \frac{2}{3}(\partial_\ell \beta^\ell ) \tilde{\gamma}_{i j }
  + (\partial_i \beta^\ell )\tilde{\gamma}_{\ell j }
  + (\partial_j \beta^\ell )\tilde{\gamma}_{\ell i }
  \nonumber\\
  &\quad
  + \beta^\ell(\partial_\ell \gamma_{ij}),
  \label{eq:gammaEvo}\\
  \partial_t\tilde{A}_{i j}
  &=
  - (D_i  D_j \alpha)^{\rm TF}e^{-4\varphi}
  - 2\alpha \tilde{A}_{im}\tilde{A}_{jn}\tilde{\gamma}^{mn}
  + \alpha K\tilde{A}_{i j }
  \nonumber\\
  &\quad
  + \alpha e^{-4\varphi}R^{\rm TF}_{ij}
  - 8\pi\alpha e^{-4\varphi}S_{i j }^{\rm TF}
  - \dfrac{2}{3}(\partial_\ell \beta^\ell)\tilde{A}_{ij}
  \nonumber\\
  &\quad
  + (\partial_i \beta^\ell )\tilde{A}_{\ell j }
  + (\partial_j \beta^\ell )\tilde{A}_{\ell i }
  + \beta^\ell (\partial_\ell \tilde{A}_{i j }),
  \label{eq:AEvo}\\
  \partial_t\tilde{\Gamma}^\ell
  &=
  - 2(\partial_i \alpha)\tilde{A}_{mj}\tilde{\gamma}^{ij}\tilde{\gamma}^{m\ell}
  + 2\alpha\tilde{\Gamma}^{\ell}{}_{mn}\tilde{A}_{ij}\tilde{\gamma}^{mi}
  \tilde{\gamma}^{nj}
  \nonumber\\
  &\quad
  + 12\alpha(\partial_i\varphi)\tilde{A}^{i\ell}
  - \dfrac{4}{3}\alpha\tilde{\gamma}^{\ell i}(\partial_i K)
  - 16\pi\alpha\tilde{\gamma}^{i\ell}J_i
  \nonumber\\
  &\quad
  + \dfrac{1}{3}(\partial_i\partial_j\beta^j)\tilde{\gamma}^{\ell i}
  + (\partial_i\partial_j\beta^\ell)\tilde{\gamma}^{ij}
  + \dfrac{2}{3}(\partial_i\beta^i)\tilde{\Gamma}^\ell
  \nonumber\\
  &\quad
  - (\partial_i\beta^\ell)\tilde{\Gamma}^i
  + \beta^i(\partial_i\tilde{\Gamma}^\ell),
  \label{eq:CGammaEvo}
\end{align}
where $R_{ij}$ in the above is defined as
\begin{align}
  R_{ij}
  &\equiv
  - \frac{1}{2}\tilde{\gamma}^{\ell m}
  (\partial_\ell\partial_m\tilde{\gamma}_{ij})
  + \dfrac{1}{2}\tilde{\gamma}_{m i}(\partial_{j}\tilde{\Gamma}^m)
  + \dfrac{1}{2}\tilde{\gamma}_{m j}(\partial_{i}\tilde{\Gamma}^m)
  \nonumber\\
  &\quad
  + \dfrac{1}{2}\tilde{\gamma}_{i\ell}\tilde{\Gamma}^\ell{}_{jm}\tilde{\Gamma}^m
  + \dfrac{1}{2}\tilde{\gamma}_{j\ell}\tilde{\Gamma}^\ell{}_{im}\tilde{\Gamma}^m
  + \tilde{\gamma}^{mn}\tilde{\gamma}_{\ell k}\tilde{\Gamma}^{\ell}{}_{nj}
  \tilde{\Gamma}^k{}_{mi}
  \nonumber\\
  &\quad
  + \tilde{\gamma}^{mn}\tilde{\gamma}_{kj}\tilde{\Gamma}^{\ell}{}_{ni}
  \tilde{\Gamma}^k{}_{\ell m }
  + \tilde{\gamma}^{mn}\tilde{\gamma}_{ki}\tilde{\Gamma}^{\ell}{}_{nj}
  \tilde{\Gamma}^k{}_{\ell m }
  \nonumber\\
  &\quad
  - 2(\tilde{D}_j\tilde{D}_i\varphi)
  - 2\tilde{\gamma}^{mn}(\tilde{D}_m\tilde{D}_n\varphi)\tilde{\gamma}_{ij}
  + 4(\tilde{D}_i\varphi)(\tilde{D}_j\varphi)
  \nonumber\\
  &\quad
  - 4\tilde{\gamma}^{mn}(\tilde{D}_m\varphi)(\tilde{D}_n\varphi)
  \tilde{\gamma}_{ij}.
\end{align}
The symbol ${}^{\mathrm{TF}}$ means the trace-free part of the value, and
$\tilde{D}_i$ is the covariant derivative associated with $\tilde{\gamma}_{ij}$.
The constraint equations are
\begin{align}
  \tilde{\mathcal{H}}
  &\equiv R
  + \frac{2}{3}K^2
  - \tilde{A}_{i j }\tilde{A}_{mn}\tilde{\gamma}^{im}\tilde{\gamma}^{jn}
  - 16\pi \rho_{\text{H}}\approx 0,
  \label{eq:Hconst}\\
  \tilde{\mathcal{M}}_i
  &\equiv
  \tilde{\gamma}^{jn}(\tilde{D}_j \tilde{A}_{ni})
  + 6(\tilde{D}_j \varphi)\tilde{\gamma}^{jn}\tilde{A}_{ni}
  - \frac{2}{3}\tilde{D}_i  K
  \nonumber\\
  &\quad
  - 8\pi J_i\approx 0,
  \label{eq:Mconst}\\
  \tilde{\mathcal{S}}
  &\equiv \text{det}(\tilde{\gamma}_{ij})-1\approx 0,
  \label{eq:Sconst}\\
  \tilde{\mathcal{A}}
  &\equiv\tilde{\gamma}^{i j }
  \tilde{A}_{i j }\approx 0,
  \label{eq:Aconst}\\
  \tilde{\mathcal{G}}^\ell
  &\equiv \tilde{\Gamma}^\ell
  - \tilde{\Gamma}^\ell {}_{i j }
  \tilde{\gamma}^{i j }\approx 0.
  \label{eq:Gconst}
\end{align}
Recently, other formulations have also been used
\cite{Brown09, Alic-BonaCasas-Bona-Rezzolla-Palenzuela12} by numerical
relativists.

\subsection{Evolution equations of matter fields
  \label{sec:matterFields}
}

For the perfect fluid, the evolution equations of matter fields
\cite{Shibata99} are given by the space--time decomposition of
$\nabla_\mu(\rho u^\mu)=0$ and $g^{\mu\lambda}(\nabla_\mu T_{\lambda \nu})=0$ as
\begin{align}
  \partial_t\rho_* &= -\partial_i(\rho_* v^i),
  \label{eq:rho_W}\\
  \partial_te_* & = -\partial_i(e_*v^i),
  \label{eq:e_W}\\
  \partial_t(\rho_* \hat{u}_i)
  &= -\partial_j\left(\rho_*\hat{u}_iv^j\right)
  - \alpha e^{6\varphi}(\partial_ip)
  - \rho_*(\partial_i\alpha) hw
  \nonumber\\
  &\quad
  + \rho_*(\partial_i \beta^n) \hat{u}_n
  - \frac{1}{2hw}\rho_*\alpha e^{-4\varphi}(\partial_i \tilde{\gamma}^{mn})
  \hat{u}_m \hat{u}_n
  \nonumber\\
  &\quad
  + 2h(w^2-1)w^{-1}\rho_*\alpha (\partial_i\varphi),
  \label{eq:rhou_W}
\end{align}
where $\rho_*\equiv\rho we^{6\varphi}$, $w\equiv \alpha u^0$,
$e_*\equiv(\rho\epsilon)^{1/\Gamma}e^{6\varphi}w$, $v^i\equiv u^i/u^0$,
$h\equiv 1+\epsilon + p/\rho$,
$\hat{u}_i \equiv he^{-4\varphi}\tilde{\gamma}_{ij}(u^0\beta^j+u^j)$, and
$\Gamma$ is a constant.
In addition, we assume $p=(\Gamma-1)\rho\epsilon$.
The relations between $(\rho_{\text{H}},J_i, S_{ij})$ and
$(\rho_*,e_*, \hat{u}_i)$ are
\begin{align}
  \rho_{\text{H}} &= e^{-6\varphi}h\rho_*w-p,
  \\
  J_i &= \rho_*\hat{u}_ie^{-6\varphi},\\
  S_{ij} &= e^{-6\varphi}w^{-1}h^{-1}\rho_*\hat{u}_i\hat{u}_j
  + p e^{4\varphi}\tilde{\gamma}_{ij}.
\end{align}
Since the four velocity $u^\mu$ is satisfied in the relation $u^\mu u_\mu = -1$,
$w$ is given as
$w=\sqrt{1+h^{-2}e^{-4\varphi}\tilde{\gamma}^{ij}\hat{u}_i\hat{u}_j}$.
In addition, for $p=(\Gamma-1)\rho\epsilon$, $h$ is given as
$h=1+\Gamma e_*\rho_*^{-1}(e_*w^{-1}e^{-6\varphi})^{\Gamma-1}$.

\section{Numerical simulations
  \label{sec:simulation}
}

In the dust case, we solve the Einstein equations and the matter evolution
equations.
With reference to \cite{Shibata99}, we set the initial data as follows.
\begin{align}
  \rho_* = a\left(1+\exp\left(\dfrac{r^2-r_0^2}{\delta r^2}\right)\right)^{-1},
\end{align}
where $a$ is the gravitational mass
$M_g = \int d^3x (\rho_{\text{H}}\psi^5 + (1/16\pi)\psi^{-7}
\hat{A}_{ij}\hat{A}_{mn}\hat{\gamma}^{im}\hat{\gamma}^{jn})$
as the unit at the initial time and $r=\sqrt{x^2+y^2+z^2}$.
This time, we set $a=4.129\times10^{-3}$, $\delta r^2=0.18M_g^2$, and $r_0=4.0M_g$.
The numerical ranges are $-9\leq x,y,z\leq 9$.
This case is static at the initial time, so we set
$X_i=\hat{A}_{ij}=J_i=\beta^i=u^i=0$.
In the dust case, we set the inner energy as $\epsilon=0$ and the pressure
$p=0$.
We assume $\hat{\gamma}_{ij} = \delta_{ij}$ and $K=0$.
Then, the extrinsic curvature $K_{ij}=0$ and $\hat{R}=0$.
We obtain the initial data $(\alpha, \gamma_{ij})$ with the above conditions by
solving Eqs. \eqref{eq:ini_psi} and \eqref{eq:maximal}.
With reference to the Schwarzschild metric, the initial step values of $\psi$
and $\alpha$ are set as $1 + 1/(1+6r)$ and $1 - 1/(3/2+r)$,
respectively.

Recently, the maximal slicing condition, Eq. \eqref{eq:maximal}, and the
minimal distortion gauge condition are almost never used in the evolution
because the numerical costs are high.
The evolution equation of $\alpha$ often uses the $1+\log$ slicing condition
\cite{Bona-Masso-Seidel-Stela95},
\begin{align}
  \partial_t\alpha = -2\alpha K.
  \label{eq:OneLogSlicing}
\end{align}
Since this condition has the characteristic of singularity avoidance, it is
widely used in the simulations for the gravitational collapse models.
We choose $\beta^i=0$.
For the rotating models of neutron stars and/or black holes,
the Gamma-driver condition
\cite{Alcubierre-Brugmann-Diener-Koppitz-Pollney-Seidel-Takahashi03} is often
used in the evolution equations of $\beta^i$.

This time, we select the grid as $\Delta x=\Delta y=\Delta z=1/60$,
$\Delta t=1/240$, and the boundary condition as the approximate asymptotic flat
boundary.
We use the fourth-order Runge--Kutta scheme with mainly the second-order
centered space difference.
However, for only the advection term, which is the first term on the right-hand
side in each of Eqs. \eqref{eq:rho_W}--\eqref{eq:rhou_W}, we use the
second-order upwind scheme.
The direction is defined by the signature of the velocity $v^i$.
\begin{figure}[htbp]
  \centering
  \includegraphics[width=0.83\hsize]{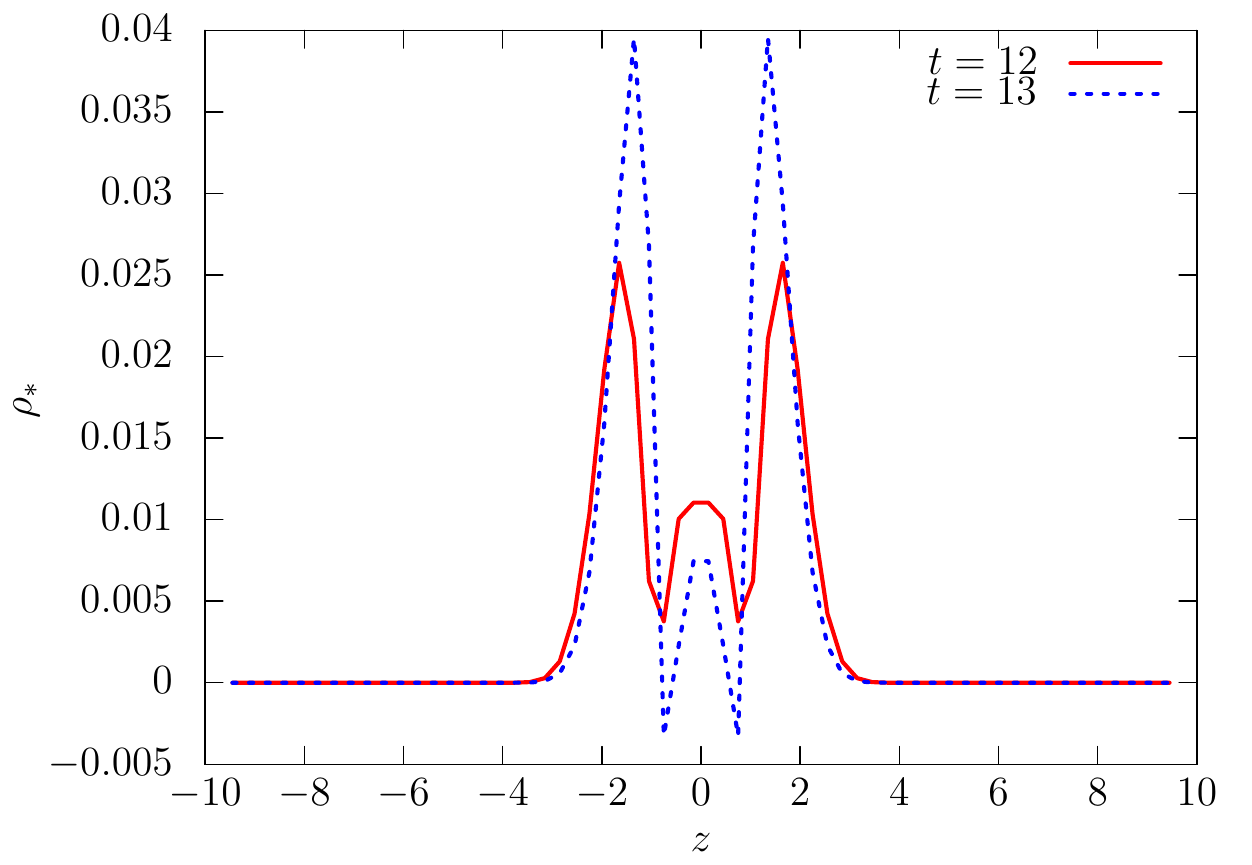}
  \caption{$\rho_*$ of $x=y=-0.45$ obtained by solving
    Eqs. \eqref{eq:varphiEvo}--\eqref{eq:CGammaEvo},
    Eqs. \eqref{eq:rho_W}--\eqref{eq:rhou_W}, and Eq. \eqref{eq:OneLogSlicing}.
    The horizontal line is $z$, the vertical line is $\rho_*$.
    The line of $t=12$ satisfies $\rho_*\geq 0$.
    On the other hand, we see that $\rho_*<0$ in $-2 < z < 2$ at $t=13$.
    \label{fig:rhostar0}
  }
\end{figure}
Fig. \ref{fig:rhostar0} shows the weighted rest mass $\rho_*$ at $t=12$ and
$t=13$ obtained by solving Eqs. \eqref{eq:varphiEvo}--\eqref{eq:CGammaEvo},
Eqs. \eqref{eq:rho_W}--\eqref{eq:rhou_W}, and Eq. \eqref{eq:OneLogSlicing}.
We see that $\rho_*<0$ at $t=13$ in the $-2<z<2$ range.
We check $\rho_*\geq 0$ during $0\leq t\leq 12$ in all the simulation ranges.
Since $\rho_*\geq 0$ is a necessary condition for successful simulations, this
simulation fails after $t=13$.

For more stable simulations, we modify Eq. \eqref{eq:rhou_W} as
\begin{align}
  \partial_t(\rho_*\hat{u}_i)
  &= [\text{Original terms}]
  + \kappa \rho_*\tilde{\gamma}^{mn}
  (\partial_m\partial_n \tilde{\mathcal{M}}_i),
  \label{eq:mod}
\end{align}
where $\kappa$ is a damping parameter.
This modification is based on the following ideas.
By using this modification, we obtain the following:
(i) the positive rest mass condition $\rho_*\geq 0$ is satisfied, (ii) the
constraints,  Eqs. \eqref{eq:Hconst}--\eqref{eq:Gconst}, are conserved, and
(iii) the total weighted rest mass
\begin{align}
  I=\int \rho_*d^3x
  \label{eq:totalMass}
\end{align}
is conserved.
The negative sign of $\kappa$ makes the simulations stable because the
dynamical equations of $\tilde{\mathcal{M}}_i$ become
\begin{align}
  \partial_t\tilde{\mathcal{M}}_i
  &= [\text{Original terms}]
  - \dfrac{8\pi\kappa\rho_*}{e^{6\varphi} w}\tilde{\gamma}^{mn}
  (\partial_m\partial_n\tilde{\mathcal{M}}_i)
  \label{eq:CPMM}
\end{align}
because of the adjusted terms of Eq. \eqref{eq:mod}.
The adjusted term of Eq. \eqref{eq:CPMM} has the dissipation effect if
$\kappa<0$ because $\rho_* /(e^{6\varphi}w)\geq 0$.
The set of Eqs. \eqref{eq:varphiEvo}--\eqref{eq:CGammaEvo}, Eqs.
\eqref{eq:rho_W}--\eqref{eq:e_W}, Eq. \eqref{eq:OneLogSlicing}, and Eq.
\eqref{eq:mod} is called as the modified system hereafter.

\begin{figure}[htbp]
  \centering
  \includegraphics[width=0.83\hsize]{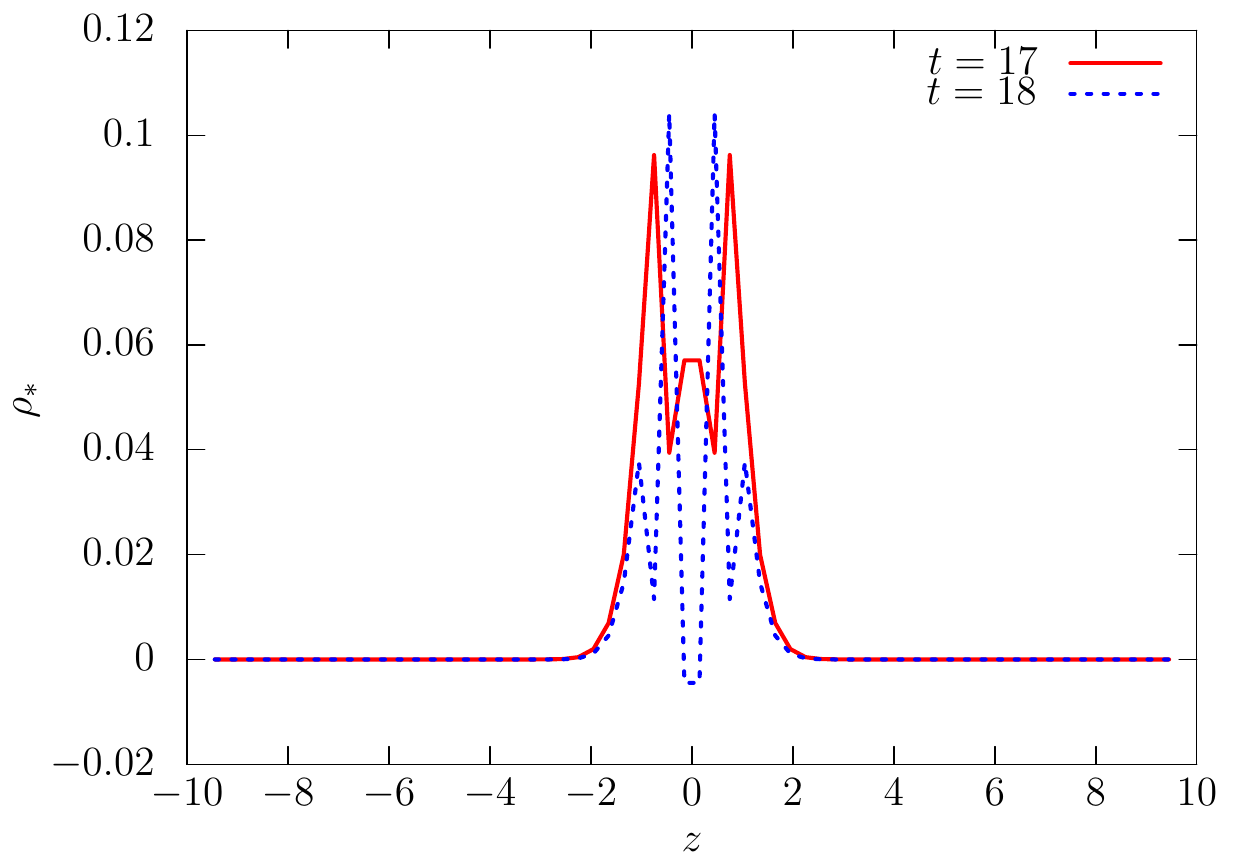}
  \caption{
    $\rho_*$ of $x=y=-0.45$ with Eq. \eqref{eq:mod} and the other conditions
    are the same as those in Fig. \ref{fig:rhostar0}.
    We set the damping parameter of Eq. \eqref{eq:mod} as $\kappa=-0.1$.
    The solid line for $t=17$ satisfies $\rho_*\geq 0$.
    On the other hand, we see $\rho_*<0$ in $-2<z<2$ at $t=18$.
    \label{fig:rhostar1}
  }
\end{figure}
\begin{figure}[htbp]
  \centering
  \includegraphics[width=0.83\hsize]{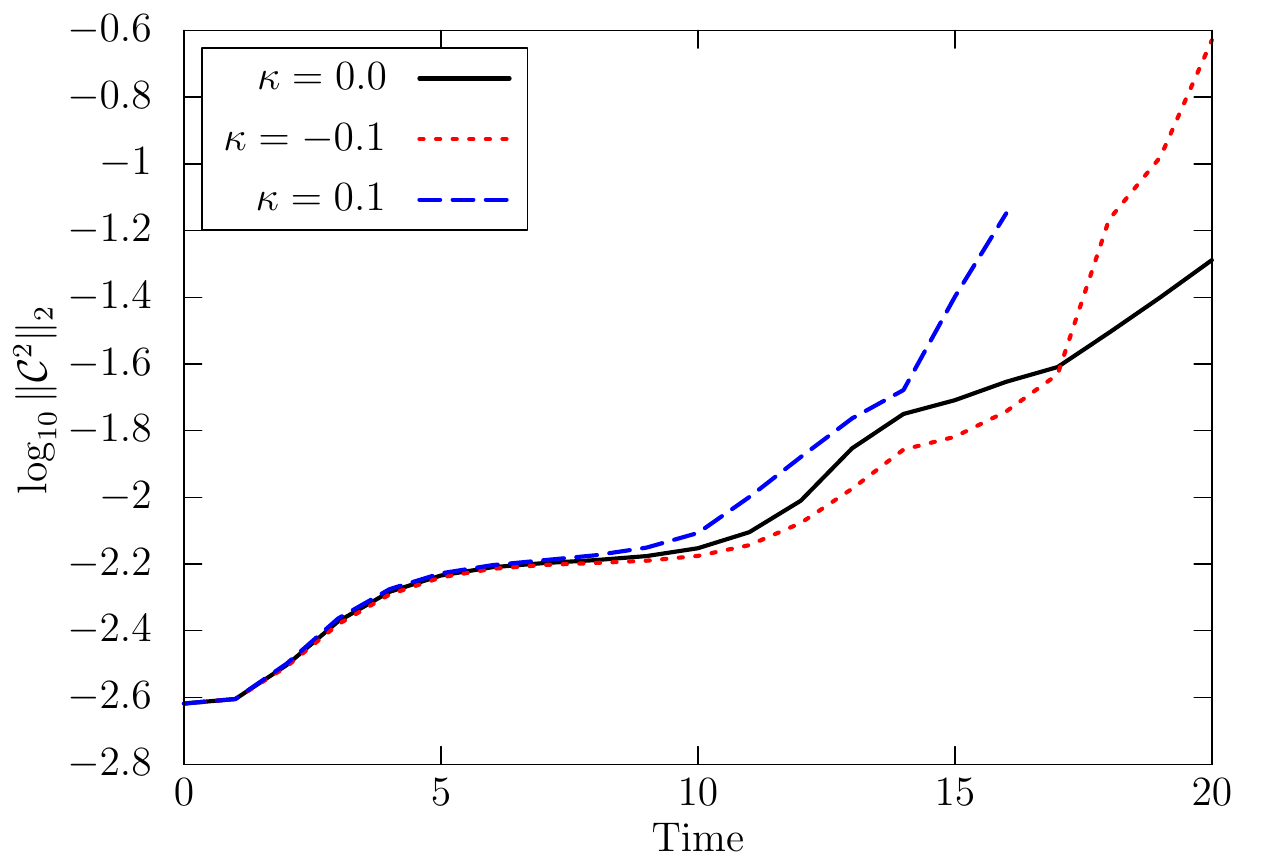}
  \caption{
    The lines show the constraint errors in the cases of $\kappa=0.0$, $-0.1$,
    and $0.1$.
    The horizontal axis is time, and the vertical axis is the logarithm of the
    L2 norm of $\mathcal{C}^2$, Eq. \eqref{eq:C2}.
    \label{fig:L2constraint}
  }
\end{figure}
\begin{figure}[htbp]
  \centering
  \includegraphics[width=0.83\hsize]{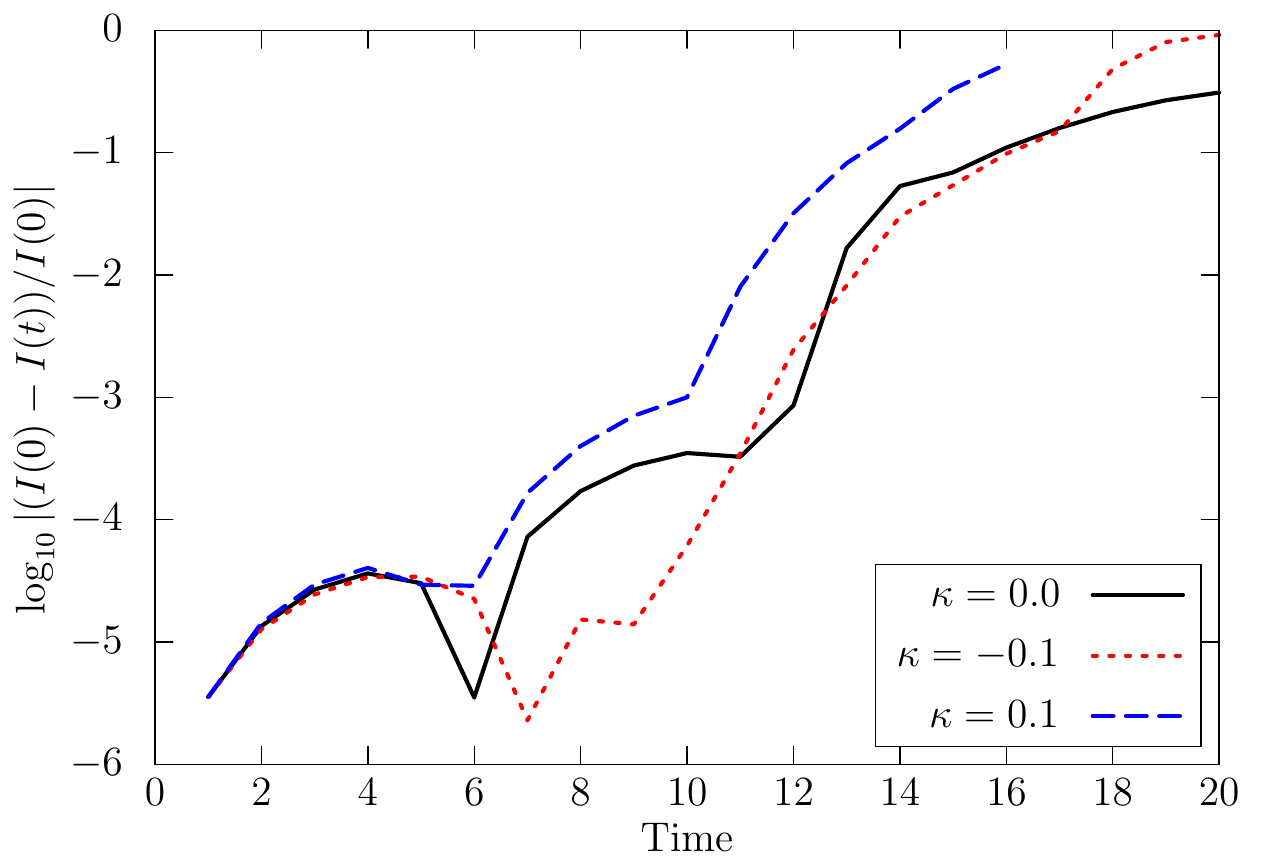}
  \caption{
    The lines show the relative errors of the total weighted rest mass $I$ in
    Eq. \eqref{eq:totalMass} against the initial values $I(0)$ for the cases of
    $\kappa=0.0$, $-0.1$, and $0.1$.
    The horizontal axis is time, and the vertical axis is the logarithm of
    $|(I(0)-I(t))/I(0)|$.
    \label{fig:Intrhostar}
  }
\end{figure}

We show the numerical results of $\rho_*$ obtained using the modified system
in Fig. \ref{fig:rhostar1}.
The numerical settings are consistent with those in Fig. \ref{fig:rhostar0}
without the evolution equation of $\rho_*\hat{u}_i$ and we set $\kappa=-0.1$.
For this simulation, we check $\rho_*\geq 0$ during $0\leq t\leq 17$ in all
simulation ranges.
In Fig. \ref{fig:L2constraint}, we show the L2 norm of the following values:
\begin{align}
  \mathcal{C}^2 = \tilde{\mathcal{H}}^2
  + \tilde{\gamma}^{ij}\tilde{\mathcal{M}}_i\tilde{\mathcal{M}}_j
  + \tilde{\gamma}_{ij}\tilde{\mathcal{G}}^i\tilde{\mathcal{G}}^j
  + \tilde{\mathcal{S}}^2 + \tilde{\mathcal{A}}^2.
  \label{eq:C2}
\end{align}
We see that the norm of the constraints $\mathcal{C}^2$ with $\kappa=-0.1$ is
less than those in other cases until $0\leq t\leq 17$.
On the other hand, the norm with $\kappa=0.1$ is larger than those in other
cases.
Thus, these results are consistent with the analytical results using Eq.
\eqref{eq:CPMM}.
Fig. \ref{fig:Intrhostar} shows the relative errors against the initial values
of the total weighted mass $I$ in Eq. \eqref{eq:totalMass}.
They are not markedly different between the cases of $\kappa=0.0$ and
$\kappa=-0.1$.

\section{Summary
  \label{sec:summary}
}

We reviewed the equations for construction of the initial values, the dynamical
equations of the Einstein equations called the BSSN formulation, and the
dynamical equations of the matter fields in the perfect fluid.
With these equations, we performed the simulations in the dust case.
We modified the system by modifying the evolution equations of the matter field,
investigated the stability analytically, and performed the simulations with the
modified system to confirm the consistency of the analytical results.
In addition, the lifetime of the simulations was extended from $t=12$ to $t=17$
with the modification.

\section*{Acknowledgments}
T.T. was partially supported by JSPS KAKENHI Grant Number 21K03354
and a Grant for Basic Science Research Projects from
The Sumitomo Foundation.
G.Y. and T.T. were partially supported by JSPS KAKENHI Grant Number
20K03740.
G.Y. was partially supported by a Waseda University Grant for Special
Research Projects 2021C-138.
%


\end{document}